# Flow-induced order-order transitions in amyloid fibril liquid crystalline tactoids

Hamed Almohammadi[1], Massimo Bagnani[1], Raffaele Mezzenga[1,2*]

[1]Department of Health Sciences and Technology, ETH Zurich, Zurich, Switzerland.
[2]Department of Materials, ETH Zurich, Zurich, Switzerland.

*Correspondence to: raffaele.mezzenga@hest.ethz.ch

Understanding and controlling the director field configuration, shape, and orientation in nematic and cholesteric liquid crystals is of fundamental importance in several branches of science. Liquid crystalline droplets, also known as tactoids, which spontaneously form by nucleation and growth within the biphasic region of the phase diagram where isotropic and nematic phases coexist, challenge our current understanding of liquid crystals under confinement, due to the influence of anisotropic surface boundaries at vanishingly small interfacial tension and are mostly studied under quiescent, quasi-equilibrium conditions. Here, we show that different classes of amyloid fibril nematic and cholesteric tactoids undergo out-of-equilibrium order-order transitions by flow-induced deformations of their shape. The tactoids align under extensional flow and undergo extreme deformation into highly elongated oblate shapes, allowing the cholesteric pitch to decrease as an inverse power law of the tactoids aspect ratio. Energy functional theory and experimental measurements are combined to rationalize the critical elongation ratio above which the director-field configuration of tactoids transforms from bipolar and uniaxial cholesteric to homogenous and to debate on the thermodynamic nature of these transitions. Our findings suggest new opportunities in designing self-assembled liquid crystalline materials where structural and dynamical properties may be tuned by non-equilibrium phase transitions.

Liquid crystalline droplets, or tactoids, form through self-organization of mesogens in concentrated aqueous suspensions[1,2] and are characterized by vanishingly low interfacial tension; when compared for example to water-oil emulsions the interfacial tension ($\gamma$) of tactoids based on filamentous colloidal is of the order of $10^{-7}$ N m$^{-1}$, that is ~100,000 times smaller than typical water-oil emulsion. While microfluidic has been widely used to study and eventually control water-oil emulsions[4-6], the hydrodynamics of liquid crystalline droplets remains poorly studied since most of the studies in these systems aimed at understanding primarily their thermodynamically equilibrium features[7]. The ability to manipulate the shape of liquid crystalline droplets by hydrodynamic forces using microfluidics allows studying their non-equilibrium phase transitions[8] and tune tactoids self-assembly structure[9], and therefore provides a promising platform to design new materials[7]. Additionally, elucidating the hydrodynamic properties of liquid crystalline tactoids can broaden our understanding of soft active matter systems exploiting anisotropic directional interactions[10-11].

Tactoids form through nucleation and growth of anisotropic domains when the mesogens are dispersed at a concentration within the isotropic (I) plus nematic (N) biphasic regime[1-3], and show various director field configurations depending on intrinsic characteristics of building blocks[1,3]. Thus, these systems are effectively a fascinating example of water-in-water emulsions with anisotropic elastic features and vanishing small interfacial tension, which endow tactoids a very unique and peculiar set of physical properties. To date, liquid crystalline tactoids have been found in dispersions of many biological rod-like systems including tobacco mosaic viruses[12], fd viruses[13], f-actin[14], carbon nanotubes[15], cellulouse[16] and more recently amyloid fibrils[1,3,17]. Amyloid fibrils, generated from common food proteins, are an appealing system due to their high potential in designing new functional materials[18,19], but also their very rich liquid crystalline behavior. As recently discovered[3,17], amyloid fibrils show transition between six different equilibrium symmetries of the nematic field configuration: homogeneous, bipolar, radial nematic, uniaxial cholesteric and radial cholesteric, with additional parabolic focal conics in bulk[17]. Such configurations result from the subtle interplay between anisotropic interfacial tension and anisotropic elastic forces in the droplet, and at equilibrium the transitions between these different classes of tactoids are accurately predicted by either scaling or variational theories[3,17]. However, far less is known under non-equilibrium conditions for these liquid crystalline droplets systems, where -as we show below- the shape of the droplets can be significantly altered by flow field with order-order transitions from one symmetry to another. Studying liquid crystalline tactoids under hydrodynamic forces opens a window on a virtually unexplored physics where anisotropic tensorial elasticity coupled with the minimal interfacial tension are anticipated to bring to light unexpected effects and to deepen the understanding of emulsions characterized by anisotropic interfacial or very low interfacial tension, such as the elusive water-in-water emulsions[20]. By using amyloid fibril tactoids as model system, here we first map flow-induced order-order transitions associated with a change in the director field configuration of the droplets during deformation. We then combine simple fluid droplet deformation theories with scaling concepts on the energy functional theory, to rationalize our experimental findings on tactoids deformation and order-order transitions, providing a general framework to study and understand the behavior of liquid crystalline droplets under deformation .

In order to deform different classes of liquid crystalline tactoids under extensional flow, we designed microfluidic chip with hyperbolic contraction zone (Fig 1a-b). An aqueous suspension of amyloid fibrils (with the mean length of $L_m$=302.9 nm and mean height of $h_m$=2.5 nm, see Fig. S1) is prepared at a concertation set within the isotropic-nematic coexistence region, allowing the formation of various classes of stable tactoids. The suspension is injected to the chip where the flow speed $U$ (controlled by syringe pump) and the length of the channel from the inlet to the contraction zone ($l_e$) are controlled to allow the solution with the formation of different classes of tactoids to equilibrate. The three main classes of tactoids that are studied here can be categorized by increase in volume (calculated as $V \approx r^2R$ by measuring major, $R$, and minor, $r$, axes of tactoids) and decrease in aspect ratio ($\alpha=R/r$) as: homogeneous nematic, bipolar nematic, and uniaxial cholesteric (Fig. 1c). The tactoids are classified based on their director field configurations[1,3]. In homogenous tactoids the director field is aligned to the long axis of the tactoid. In the case of bipolar tactoids, the director field always follows the interface. The uniaxial cholesteric configuration is characterized by its typical striped texture and quantified by pitch value ($P$), which is two times the band-to-band distance. The configurations of the tactoids at rest are directly linked to their confined structures and the transitions between different classes of tactoids at equilibrium are theoretically predicted by Nystrom *et al.*[3] as a function of $V$ and $\alpha$; the pitch is found to correlate with the shape of the cholesteric droplet where upon an increase in $V/\alpha$, the pitch decreases asymptotically[3].

The tactoids, when passing through the hyperbolic geometry, experience extensional flow[21]. The width of the contraction zone, $w(x)$, with length ($l_c$), height ($h$), upstream width ($w_u$), and throat width ($w_t$), is shaped according to the function proposed in Ref. 21 as $w(x)=x_1/(x_2+x)$, where $x_1=l_c w_u w_t/[2(w_u-w_t)]$ and $x_2=l_c w_t/(w_u-w_t)$. Accordingly, for a given volumetric flow rate ($Q$) and with the assumption that shear flow induced by bounding wall is negligible, the flow speed, $u_x=Q/[2w(x)h]$, increases linearly as $x$ increases, leading to a constant extension rate, $\dot{\varepsilon}_{xx}=\frac{\partial u_x}{\partial x}=Q/(2x_1h)$, along the hyperbolic contraction zone ($0 \leq x \leq l_c$), Fig. 1d. To ensure that the assumption of negligible shear flow effect by the channel wall holds valid in the analysis, only the tactoids passing close to the centerline of the channel ($y = 0$) were analyzed.

Various dimensions of contraction geometry (see Methods) and flow rates ($0.3 \leq Q \leq 1.8$ mm$^3$ h$^{-1}$) are used to induce different extensional rate $\dot{\varepsilon}_{xx}$, i.e. $0.004 \leq \dot{\varepsilon}_{xx} \leq 0.020$ s$^{-1}$.

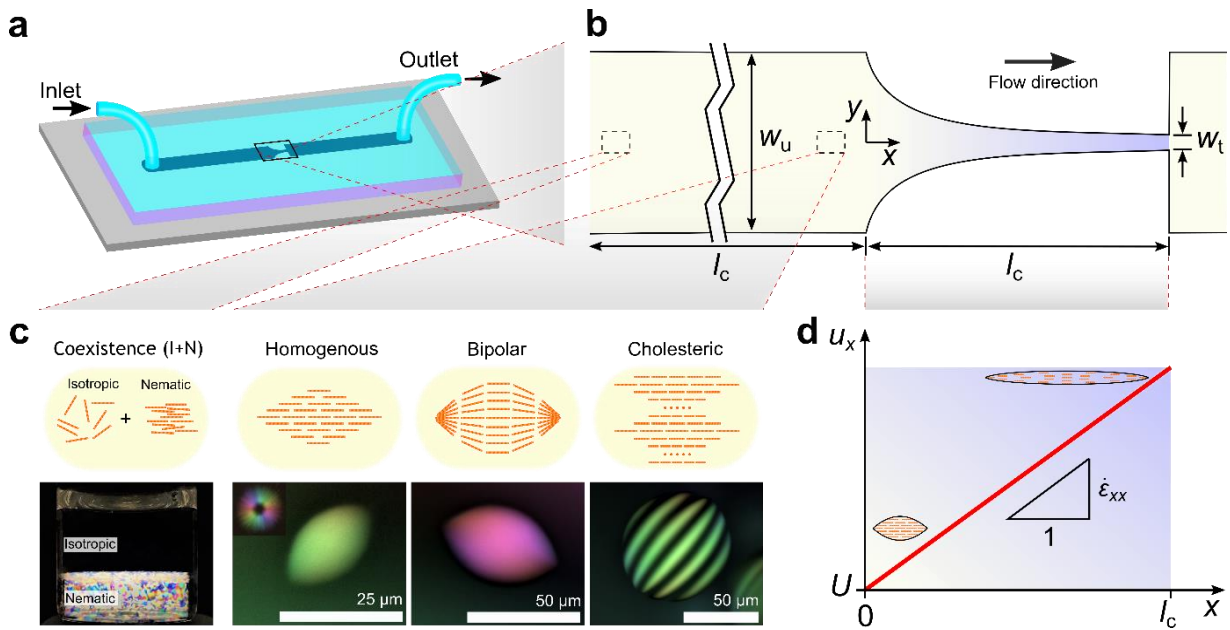

**Fig. 1 | Microfluidic chip to expose different classes of tactoids to extensional flow. a** Schematic of the microfluidic chip. **b** The hyperbolic contraction geometry. The origin of the coordinate system is located at the contraction inlet with $x$ axis along the channel centerline. **c** A suspension of amyloid fibrils at concertation in the I+N coexistence region is injected to the microfluidic chip. LC PolScope images of the three classes of

homogenous, bipolar, and cholesteric tactoids studied. **d** Schematic plot of the average velocity of flow, that varies linearly along the contraction zone to allow a constant extension rate.

When injected in the microfluidic chip, the tactoids align their long axes parallel to the flow direction while approaching the contraction zone ($x < 0$). In the contraction zone ($0 < x < l_c$), the tactoids are subjected to extensional flow. As shown in Fig. 2, a multitude of complex events can be resolved by this approach. The most common event is the continuous stretch of homogenous tactoids while maintaining their director symmetry, independently of the extension; in contrast, the bipolar and cholesteric tactoids undergo order-order transition to homogenous tactoids (Fig. 2a-c) at a critical extension. To distinguish different classes of tactoids, we examined tactoids between crossed polarizers and tracked the texture of tactoids during the elongation process. In the Supplementary Information we further show how displaying different textures of the tactoids at different angles relative to crossed polarizers, allows resolving unambiguously the deformation-induced order-order transitions of bipolar and cholesteric tactoids into homogenous tactoids. Right after the contraction ($l_c < x \lesssim 2l_c$) the tactoids rotate by ~90°, aligning their long axes perpendicular to flow direction, recovering their original starting phases. Such a parallel and perpendicular alignment of tactoids to the flow direction before and after the contraction zone, respectively, was shown previously for anisotropic cylindrical and disklike colloidal particles[22]; however, we show here that similar behavior can be observed also in tactoids, extending the validity of the argument to larger length scales. Additional events include break-up and coalescence of the tactoids. Example of the break-up of a bipolar tactoid into a homogenous and a bipolar tactoid is shown is Fig. 2d, reminiscent of the liquid filament break-up as a function of its aspect ratio and rheological properties[23,24]. Coalescence of tactoids into larger volume can also occur when two tactoids approach simultaneously the contraction zone, as shown in Fig. 2e. Such a break-up and coalescence processes of tactoids demonstrate that extensional fluid flow can be used to manipulate tactoids volume by disentangling it from the mass transport between continuous and dispersed phase, which typically occurrs during nucleation and growth. This is of critical importance in designing the systems with desired classes of tactoids. Note that the coalescence events shown here occur due to hydrodynamic focusing region of the selected geometry where the tactoids come together following the flow streamlines.

A distinctive tract of the undergoing deformation mechanisms is the very low interfacial tension characterizing the tactoids. To highlight this further, control experiments with simple fluids (oil droplet in water-glycerol mixture) with interfacial tension in the order of ~ 0.01 N m$^{-1}$ were performed and compared with those on the tactoids (Fig. 2f). To make a meaningful comparison, the relevant parameters in droplet deformation[25], i.e. droplet size, the viscosity of the droplet, viscosity of the medium, and accordingly the viscosity ratio of droplet to medium, as well as extension rate, were kept virtually identical (see Supplementary Information for more details). Note that, although the viscosities of the liquid crystalline phases vary depending on the shear rate (see SI), for the sake of comparison and following the common assumption in the context of the droplet deformation[25], zero-shear viscosity values of liquid crystalline phases were considered here. The extension rate and droplet size are set to be equal to those of the corresponding cholesteric tactoid. When comparing Fig. 2c and 2f, it becomes clear that, while the tactoid reaches a length ratio of ~10 upon deformation, the oil droplet remains almost undeformed, highlighting the remarkable effects of very low interfacial tension of the tactoids.

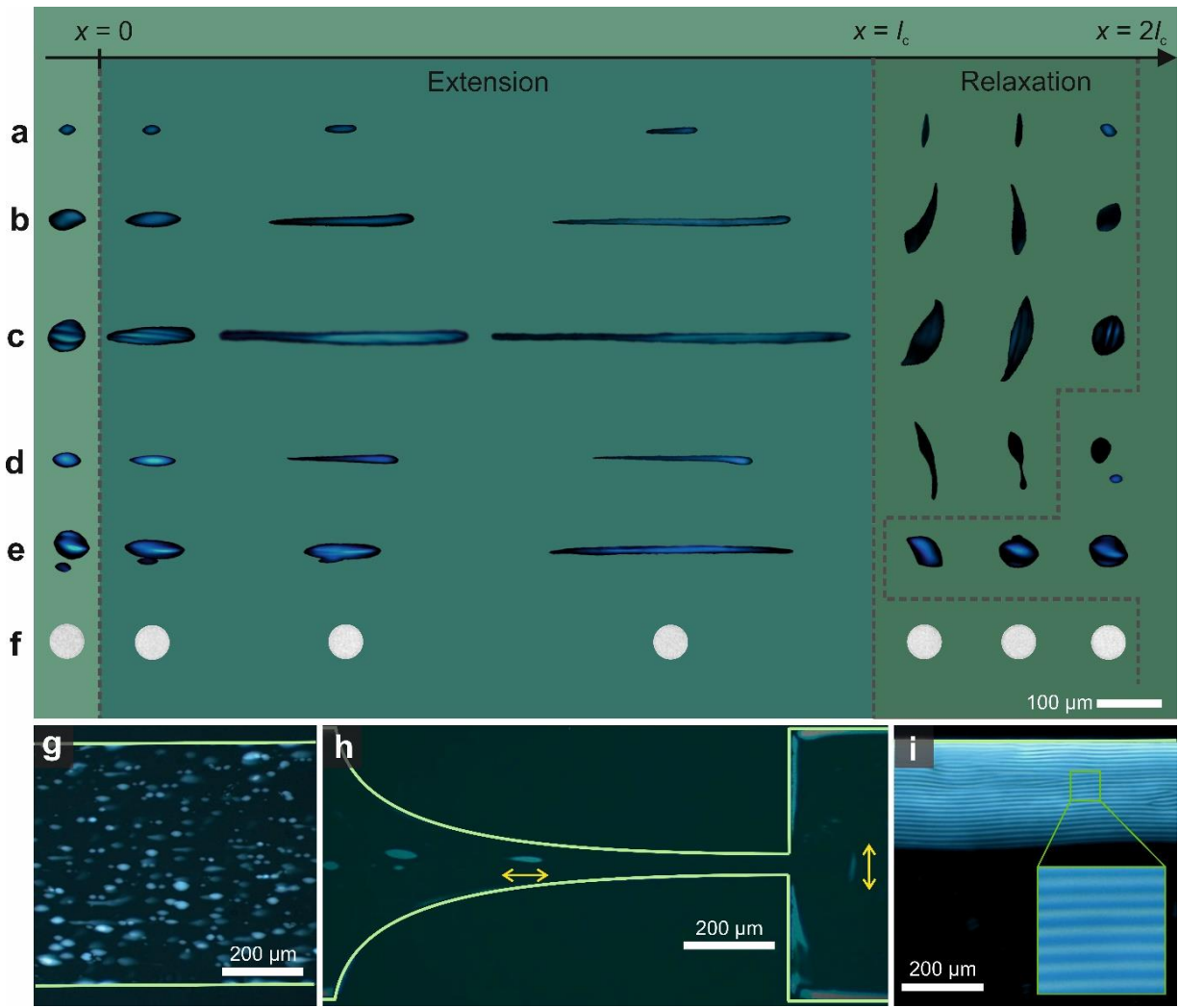

**Fig. 2 | Fate of amyloid fibril tactoids under extensional flow field. a-c** The tactoids pass through the contraction zone (flow direction is from left to right) while having their long axes aligned with the flow direction. In the contraction zone ($0 < x < l_c$), the homogenous tactoids maintain their original phase (**a**), while the bipolar (**b**) and cholesteric (**c**) tactoids undergo order-order transitions to homogenous tactoids. After the contraction ($l_c < x \lesssim 2l_c$) the tactoids rotate ~90° and recover their original phase. **d** Break-up of a bipolar tactoid. **e** Coalescence of two tactoids resulting into a tactoid with larger volume. **f** A control experiment on oil-in-water droplets as observed by bright field images, reveals essentially undeformed oil droplet through the same flow field conditions. **g** Alignment of the tactoids in flow direction. **h** Rotation by ~90° of the tactoids exiting the channel. **i** Formation of uniform bulk cholesteric phase as large as 200 μm with few disclination lines (the figure shows half of the channel) close to exit of the channel, appeared by flowing the suspension in the chip for more than two days.

The measurements of the deformations of tactoids under extensional flow are shown in Fig. 3a-c. In order to quantify the deviation of the shape of tactoid from its initial shape we use the Taylor deformation parameter, defined as[26]: $\mathcal{D}=(R-r)/(R+r)$, having zero value for a sphere and close to unity when the tactoid reaches extreme deformation, such as a long liquid filament. A linear trend is observed for the tactoids deformation parameter as a function of the adimensional time, $\tau = t\dot{\varepsilon}_{xx}$, with $t$ the time. Similar behavior is also reported for simpler viscoelastic fluids, when the stretching timescale is substantially lower than the time required for a droplet to reach steady state deformation[27,28]. Thus, time-dependent deformation of tactoids follow $\mathcal{D}=(\partial\mathcal{D}/\partial\tau)\tau+\mathcal{D}_0$, where $\mathcal{D}_0$ captures the initial shape of the tactoids. In the Supplementary Information, we provide arguments allowing the estimation of $\mathcal{D}_0$ based on the equilibrium shape of the tactoids. As depicted in Figure 3a-c the transition between various classes of liquid crystalline tactoids is controlled by elongation, with bipolar and cholesteric tactoids turning into homogenous tactoids at a critical threshold. For the cholesteric to homogenous transition, there is a range of the deformation where the tactoids do not fall into a well-defined class, for example when the striped texture exists only partially along the tactoid. We indicate such a transient range with empty symbols in Fig. 3c.

Assuming that the tactoid geometry is axisymmetric with respect to its long axis during the elongation, the volume of the tactoids is measured and found to be almost constant (Fig. 3d-e), confirming that the tactoids undergo uniaxial extension during the deformation. However, there is slight decrease in the calculated volume for the cholesteric droplets during the deformation, which suggests that for this system the shape is not entirely axisymmetric.

To rationalize the effect of hydrodynamic forces on the elongation of the tactoids, we examined the deformation of the tactoids at different Capillary numbers, $Ca=\mu_I R_{eq}\dot{\varepsilon}_{xx}/\gamma$, showing the ratio of viscous stretching forces to the surface forces that resists the stretching, where $\mu_I$ is the viscosity of the continuous phase and $R_{eq}$ is the equivalent radius of tactoids defined as:[29] $(r^2R)^{1/3}$, Fig. 3g-i. We measured $\mu_I$ to be 0.121 Pa s and $\gamma$ is calculated to be $1.1 \times 10^{-6}$ N m$^{-1}$, see Supplementary Information for more details. Inertial forces are neglected here as Reynolds number, $Re=\rho_I R_{eq}U/\mu_I$, expressing the ratio of inertia to viscous forces, with $\rho_I$ is the viscosity of the continuous phase, is very low and in the order of ~$10^{-7}$ (see Supplementary Information for details). In the experiments, we kept the flow velocity lower than a critical value over which the fibrils in the isotropic phase aligns. Above such a flow velocity, that is well predicted by De Gennes[30,31], the medium becomes nematic and the induced birefringence of the medium blends with the one by the tactoids, making the tactoids undetectable. The continuous linear stretching behavior is observed at different *Ca* numbers for deformation parameter as a function of dimensionless time, i.e. $\mathcal{D} \sim \tau$. Such a continuous stretching and extreme deformation of the tactoids conceptually show that the very low interfacial and elastic forces of tactoids are not sufficient to balance the hydrodynamic force and keep the shape of the tactoids steady[32]. Having the relation of the $\partial\mathcal{D}/\partial\tau$ as a function of *Ca* number together with $\mathcal{D}=(\partial\mathcal{D}/\partial\tau)\tau+ \mathcal{D}_0$, from earlier discussion, we obtain the following simple expression describing the tactoids time-dependent deformation:

$$\mathcal{D} = m Ca\tau + \mathcal{D}_0 \qquad (1)$$

where $m$ is the slope of the dashed lines in Fig. 3g-i. Having $\mathcal{D}$ value from Eq. 1, one can get the aspect ratio of the tactoids as $\alpha=(1+\mathcal{D})/(1-\mathcal{D})$. Equation 1, that predicts the continue deformation of the tactoids as a function of extension rate and the restoring effect of the interfacial tension, agrees well with the linear deformation concept considered for simple fluids when the stretching timescale of the fluid droplets is significantly lower than their characteristic relaxation time[33-35]. Note that, although we assumed the same viscosity for all classes of tactoids, the viscosity is expected to decrease when going from cholesteric and bipolar to homogenous tactoids since the rods within the tactoids are more oriented toward the stretching direction in homogenous tactoids. Essentially, as it is also reflected in shear thinning behavior of the liquid crystals (see SI), the viscosity of the liquid crystals decreases as the fibrils get oriented in flow direction[36]. Such a decrease in viscosity can explain the increase in $m$ value (i.e. resulting in larger deformation, Fig. 3g-i) for the progression cholesteric → bipolar → homogenous.

It is important to note that Eq. 1 can be extended to other materials based liquid crystalline tactoids that have different elasticity from that of amyloid fibrils, as it has been found that elasticity does not have significant effect on the trend of the droplet deformation[37].

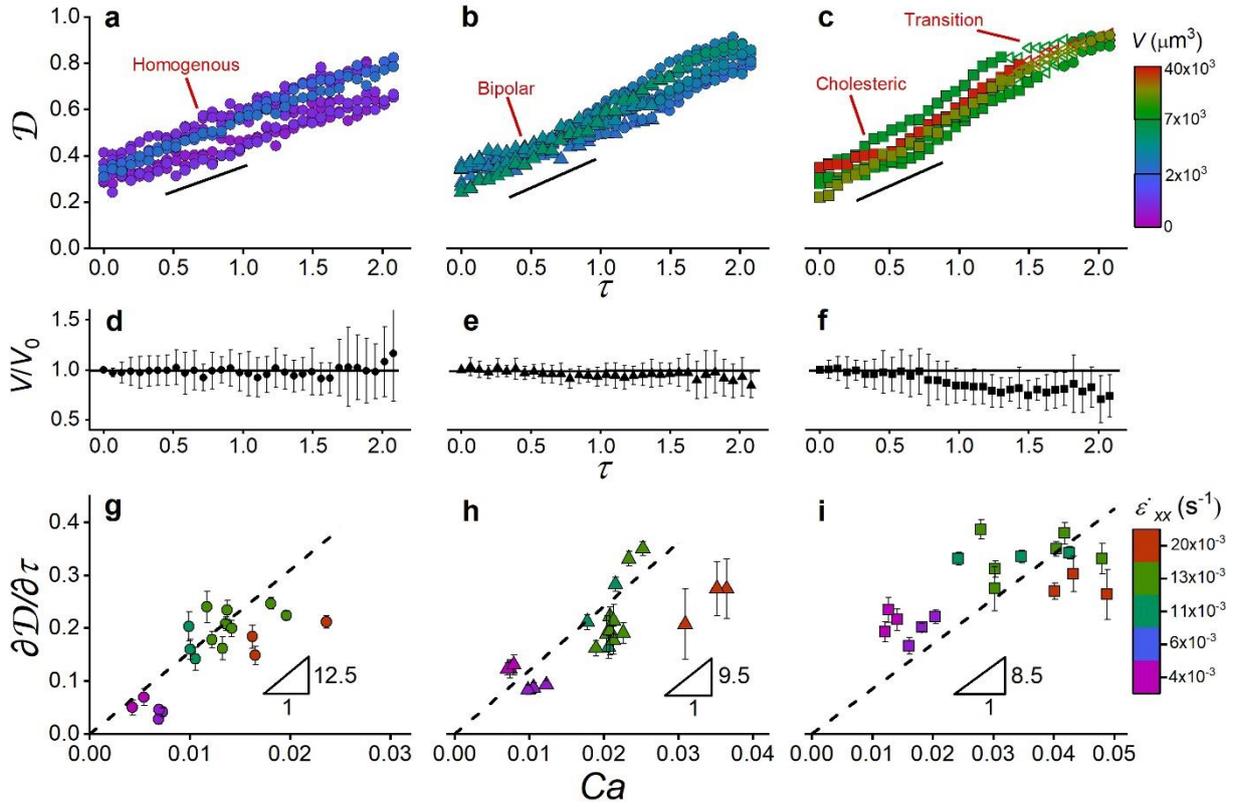

**Fig. 3 | Deformation of different classes of liquid crystalline tactoids.** The data points presented by filled circle, triangle, and square correspond to homogenous, bipolar, and cholesteric tactoids, respectively. The empty symbols denote the transition regime of cholesteric to homogenous tactoids. **a to c,** deformation of tactoids under extension rate of 0.013 s$^{-1}$ that shows linear change in deformation parameter $\mathcal{D}$ as a function of $\tau$ and order-order transitions of the bipolar and cholesteric tactoids to homogenous ones. **d-f** variation in volume of the tactoids under extension rate of 0.013 s$^{-1}$ showing that the volume of the tactoids remain essentially constant. **g-i** the linear dependence of $\mathcal{D}$ on $\tau$ are tested for different $Ca$ numbers. A linear fitting is used as suggested by simple fluids theories. A total of 59 tactoids has been analyzed. The error bars denote standard deviation.

Having quantified the deformation of tactoids under extensional flow by Eq. 1, we now rationalize the director field transitions of liquid crystalline tactoids under deformation by looking at the free-energy landscape of the tactoids using a scaling form of Frank–Oseen elasticity theory. According to this theory, there are two energetic contributes associated with the tactoid at equilibrium, namely the bulk elastic and surface energies, and the total free energy of the tactoid can be expressed according to the following scaling form[3]:

$$U \sim \frac{Kr^2}{R} + \gamma Rr\left[1 + \omega(r/R)^2\right] + \frac{1}{2}K_2(\theta + q_\infty)^2 r^2 R \qquad (2)$$

where the first and last term account for bulk elastic energy and the middle term describes the surface energy of the system. The importance of each term progressively shifts from left to right as the tactoids grow. Splay and bending energy are embedded in the first term, where $K$ is the Frank elastic constant for splay and bending that are assumed to be equal here. The twist elastic energy is given by third term, where $K_2$ represents Frank twist elastic constant and $q_\infty$ is the chiral wave number, equal to $2\pi/P_\infty$ with $P_\infty$ the natural pitch of the system. Finally, the middle term in Eq. 2, is the surface energy of the system that accounts for both interfacial tension and anchoring strength, $\omega$. We measured and theoretically calculated the main constant entering in the energy functional to be: $K=1.0 \times 10^{-11}$ N, $K_2 = 2.0 \times 10^{-12}$ N, $P_\infty = 25.6$ μm, and $\omega = 1.17$ (see Supplementary Information for the details).

To determine the critical aspect ratio at which the order-order transitions occurs for a tactoid with a given volume, we examine the interplay between the terms in Eq. 2 upon transitions. The bipolar to homogenous transition is associated with the entire conversion of the bulk energy of the bipolar tactoids, $\frac{Kr^2}{R}$, into the anchoring part of the surface energy of the homogeneous tactoids, $\gamma R r \omega (r/R)^2$, for which the homogeneous director field imposes zero elastic energy. Thus, by energy conservation, at the transition $\frac{Kr^2}{R} \approx \gamma R r \omega (r/R)^2$ must hold, which can be reworked into:

$$\alpha|_{\text{bipolar} \to \text{homogenous}} \approx V \left(\frac{\gamma \omega}{K}\right)^3 \quad (3)$$

Using the similar approach, when the cholesteric to homogenous transition happens, the bulk energy increases by $\frac{1}{2} K_2 q_\infty^2 r^2 R$ as θ changes from $-q_\infty$ to 0 in Eq. 2, while the surface energy decrease by $C_i \gamma R r \omega (r/R)^2$. The term $C_i$ is a constant that accounts for the change in surface anchoring energy when cholesteric changes to homogenous tactoids as the rods are anchored to surface in different fashions in two systems. Overall, from energy conservation, we have $C_i \gamma R r \omega (r/R)^2 \approx \frac{1}{2} K_2 q_\infty^2 r^2 R$, which upon simplifying and omitting pre-factors gives:

$$\alpha|_{\text{cholesteric} \to \text{homogenous}} \approx V^{-1/5} \left(\frac{\gamma \omega}{K_2 q_\infty^2}\right)^{3/5} \quad (4)$$

Since Eq. 3 and Eq. 4 are derived from a scaling form of the energy functional, they are correct within a prefactor. Yet, as can be seen by Fig. 4a, which brings together the scaling arguments and the experimental data providing the phase diagram of liquid crystalline tactoids at different elongation ratios, these scaling expressions capture very well the experimental findings and the corresponding order-order transitions. We argue here that as long as the elongation of the tactoids is uniaxial, one can take the theoretical framework presented in Fig. 4a as a universal phase diagram to predict the order-order transition of the tactoids under any type of the external force field.

To further elucidate the cholesteric droplet deformations and director field transitions, we measured the cholesteric pitch value under constant extension rate of 0.013 s$^{-1}$ as shown in Fig. 4b. Two main features can be noted from Fig. 4b: (i) the order-order transition of uniaxial cholesteric to homogenous tactoids that is observed by the decrease and annihilation of the cholesteric pitch and (ii) the number of cholesteric bands remains almost constant during the elongation, allowing us to state that that the pitch and short axis of the cholesteric tactoids evolve proportionally, i.e. $P/2 \sim r$ (see the inset in Fig. 4c). This, together with aforementioned discussion indicating that the volume of the droplet remains constant during elongation, i.e. $V \sim r^3 \alpha = constant$, giving $r \sim \alpha^{-1/3}$, allows us concluding that $P/2 \sim \alpha^{-1/3}$. Figure 4c presents the experimental data of the evolution of the pitch as a function of deformation, showing a good agreement with the expected scaling law. Such a change in cholesteric pitch value, for instance from $P/2=15$ µm to $P/2=6$ µm for a tactoid of $V=7500$ µm$^3$, demonstrates the possibility of tuning the cholesteric pitch and hence the wavelength of light that is transmitted/reflected by them[38]. Figure 4c, does not allow to draw conclusions on the pitch evolution beyond $P/2=6$ µm (range highlighted in gray); however it is inferred that in order to go from $P/2=6$ µm to infinite, as should be for homogeneous tactoids, a sudden jump must be expected, which would point to a first order thermodynamic transition. While this cannot be conclusively assessed here, to rule out possibility of a possible but not observable pitch beyond $P/2=6$ µm in Fig. 4c, we checked the images resolution, and found that being the resolution around six times finer than the minimum observed pitch value, a continuous evolution of the pitch beyond $P/2=6$ µm can be ruled out, which re-inforce the expected first order nature of the transition. Additionally, the deformed cholesteric tactoids that have undergone an order-order transition to homogenous tactoids in the contraction zone, were checked at the beginning of the relaxation (right after contraction zone, see Fig. 2c) and no sign of pitch was found. This, together with the smooth evolution of the pitch in the observable window (Fig. 4b) further discard the possibility of a continuous decrease of pitch beyond $P/2=6$ µm, reinforcing the argument on the first-order thermodynamic nature of this transition.

To conclude, we have shown that using microfluidic to produce a purely extensional flow can be used as a mean to induce alignment, deformation, order-order transitions, coalescence and break-up of amyloid-based liquid crystalline tactoids. Under such an imposed extensional flow, the tactoids are shown to undergo extreme deformations at a Capillary number as low as $\sim 10^{-2}$. By combining theory and experiments, we have been able to rationalize the threshold at which order-order transitions are expected and to debate on the thermodynamic nature of these transitions. Our results open a new window on non-equilibrium features of colloidal systems at extremely small interfacial tension and can pave the way to new strategies in the design of self-assembled complex fluids.

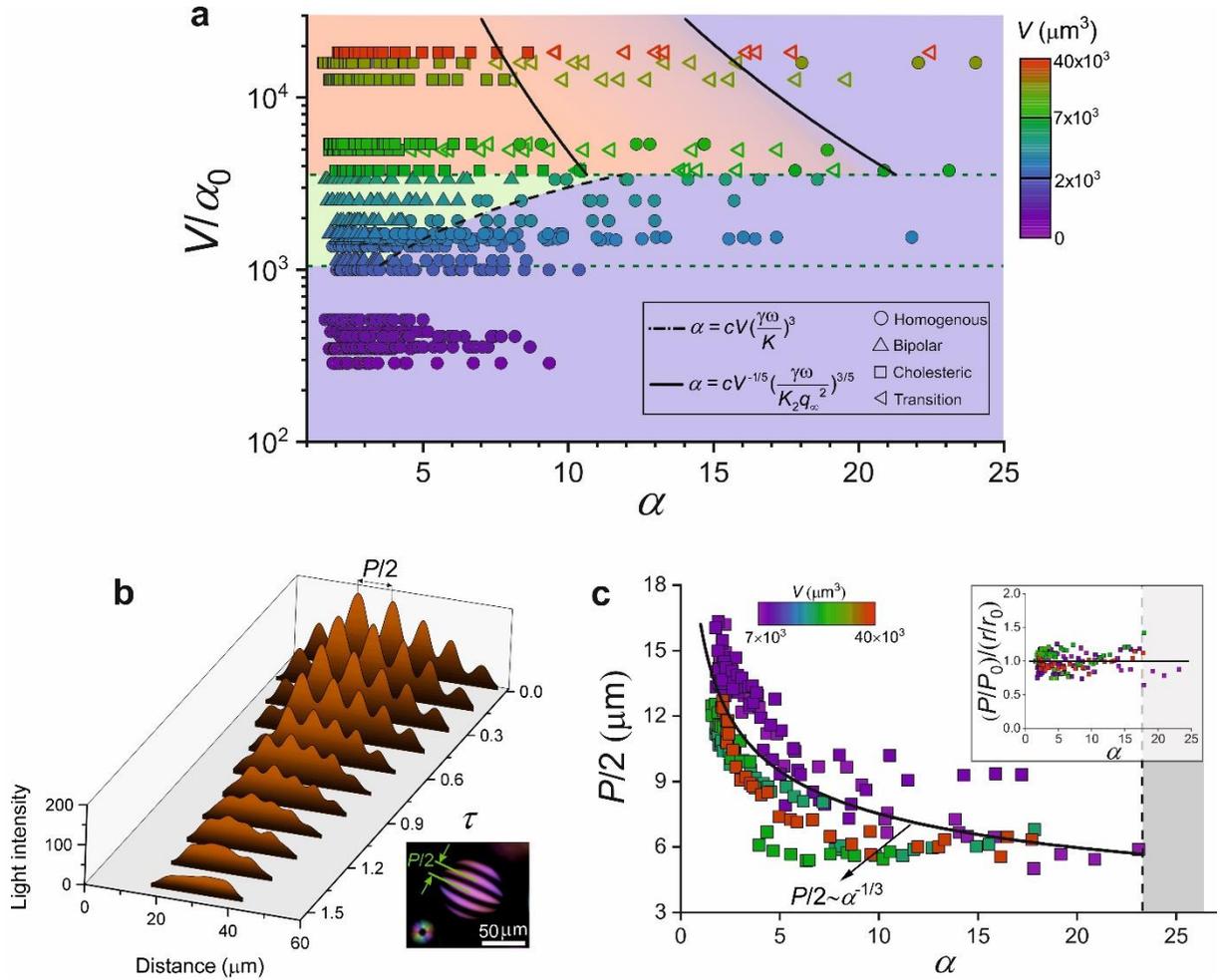

**Fig. 4 | Phase diagram of nematic–cholesteric tactoids undergoing order-order transitions induced by an extensional flow and evaluation of cholesteric pitch at different elongation ratios. a** filled circle, triangle, and square symbols denote homogenous, bipolar, and cholesteric tactoids. The empty symbols correspond to transition regime of cholesteric to homogenous tactoids. The developed theory predict the transition of the bipolar to homogenous (dashed line) and cholesteric to homogenous (solid lines) tactoids. The constant $c$ value is 0.8 for bipolar to homogenous transition; and for cholesteric to homogenous tactoids, the $c$ value is 14.5 for lower and 29.0 for upper boundary of transition region. **b** evaluation of the cholesteric pitch under constant extension rate of 0.013 s$^{-1}$ showing both decrease of cholesteric pitch and order-order transition of the cholesteric tactoids. **c** change in cholesteric pitch as a function of elongation ratio that scales as $P/2 \sim \alpha^{-1/3}$. The inset shows that ratio of $P/r$, for a given cholesteric tactoid, remains essentially constant at different elongation ratios.

**Methods**

**Preparation of amyloid fibrils suspension.** First, we dispersed 6 grams of β-lactoglobulin, purified from whey protein (see Ref. 39 for details), in 300 ml Milli-Q water that were set to pH 2 by adding HCl prior to dispersion, giving 2 w/w of β-lactoglobulin/Milli-Q water. This was followed by filtration of the dispersion through 0.45 µm Nylon syringe filter (Huberlab) to remove possible aggregation. Then, following heat-induced denaturation process, we let the solution over a hot plate (IKA, RCT basic) for 5 hrs at 90 °C, while the magnetic stirrer (length: 3cm and diameter: 0.6 cm) was rotating at ~1000 rpm inside the suspension to avoid forming of any gel close to air interface.

The amyloid fibrils in the suspension were cut by applying mechanical shear force, giving the length distribution reported in the Supplementary Information. To clear the suspension from any unreacted monomers and peptides, it was dialyzed for 5 days using 100 kDa (MWCO) Spectra/Por dialysis membrane (Biotech CE Tubing) against 10L Milli-Q water, adjusted at pH 2, with everyday bath change. The desired concentration for the suspension was adjusted by up-concentrating it through reverse osmosis using 6-8 kDa (MWCO) Spectra/Por 1 dialysis membrane (Standard RC Tubing) against 10 wt% polyethylene glycol solution (mol wt: $M_r$~35000, Sigma Aldrich) that was set to pH 2. The end concentration used in the experiments was 2.2 wt%, that is in the biphasic region, with 2.0 wt% and 2.5 wt% concentrations for isotropic and nematic phases, respectively, that were measured after macroscopic phase separation occurred. Note that, after every step, the suspension was checked between cross polarizer for any gelation or larger protein aggregates. The concentration measurement was made gravimetrically using Mettler AT20 microbalance.

**Characterization of the physical properties of the amyloid fibrils.** A suspension of 0.01 wt% was prepared from a 2.2% of amyloid fibrils suspension used in the experiments by diluting it in pH 2 Milli-Q water. As described elsewhere[14], to perform the AFM measurement, the amyloid fibrils solution was deposited on a freshly cleaved mica by placing a droplet of prepared suspension for 120 s. This is followed by rinsing the mica with Milli-Q water and drying it with pressurized air. MultiMode VIII scanning probe microscope (Bruker) was used to acquire the images, while running in tapping mode at ambient conditions. The images acquired in AFM were analyzed by FiberApp[40] allowing to track each fiber and measure the length and height distribution, repsectively.

**Microfluidic.** To fabricate the microfluidic chip, the standard soft lithography procedure was followed[41]. The microfluidic channel was made by plasma-bonding of the PDMS channel to plain glass slides (Corning 2947). The attached PDMS channel was made from a 10:1 mixture of polydimethylsiloxane (PDMS) monomer and curing agent (Dow Corning Slygard 184).

The used microfluidic chip had rectangular cross section. The four sets of geometrical dimensions used in this study are: 1) $w_u$ = 600 μm, $w_t$ = 50 μm, and $h$ = 100 μm; 2) $w_u$ = 1000 μm, $w_t$ = 200 μm, and $h$ = 200 μm; 3) $w_u$ = 1000 μm, $w_t$ = 150 μm, and $h$ = 200 μm; 4) $w_u$ = 1000 μm, $w_t$ = 100 μm, and $h$ = 200 μm. The length of the contraction zone was kept the same for all geometrical dimensions at $l_c$ = 1000 μm, while different entrance lengths were used, 1500, 3000 and 4500 μm, respectively.

**Sample characterization.** To identify the symmetry of the tacoids, polarized optical microscopy Zeiss with attached camera (AxioCam MRc) was used. Unless otherwise noted, objectives 5× (Achrostigmat) and 10× (Plan Neofluar) were used to capture the images between crossed polarizers. For the cases of under a flow field measurement, time series acquisition mode at frame at rate of 12 frame per minutes were used. The acquired images were analyzed using ImageJ. For quantitative analysis of tactoid defromation, the microfluidic channel was mounted on the microscope keeping the flow direction at 45° with respect to one of crossed polarizers, while to support the order-order transition analysis the microfluidic channel was rotated at different angles with respect to crossed polarizers, see Supplementary Information for more details.

**Experimental detail.** All of the tests were run at room temperature. The suspension was filled in a 250 μl syringe (Hamilton) and pumped at different flow rate using syringe pump (Harvard Apparatus). For the external fluidic interconnects, short pieces of flexible tubing with inner diameter 0.8 mm was used. To connect the tubing to the microfluidic channel, the needle with inner and outer diameters of 0.34 mm and 0.64 mm, respectively, was used. A new microfluidic chip was used upon significant deposition of the suspension on the channel wall to avoid any disturbance on the flow field.

**Rheological measurements.** To measure the viscosity of the liquid crystalline phases, a MCR 502 (Anton Paar) rheometer with PP25 parallel plate geometry and 1 mm gap, was used. Due to possible errors associated to the measurement of low viscosities in this geometry, the validity of the measurement was checked using mixture of water-glycerol with known viscosity that was in the same range of liquid crystal phases viscosities. All of the measurements were completed at room temperature, similar to the conditions at which the experiments are performed. To avoid possible evaporation of the suspension during the measurement, a solvent trap was used in all the measurements.

**Acknowledgements** We thank J. Adamcik for atomic force microscopy measurements, X. Cao (ETHZ) for assistance in fabrication of microfluidic chips, and P. Fischer (ETHZ), P. Azzari, and M. Diener for valuable discussions. Prof. Andrew de Mello (ETHZ) is kindly acknowledged for granting access to his laboratory to fabricate microfluidic chips.

**Author Contributions** H.A. designed and performed the experiments, contributed the theoretical scaling, analyzed the data, interpreted the results, carried out the theory, and wrote the manuscript. M.A. contributed to the experiments. R.M. designed and directed the research, analyzed the data, interpreted the results, developed the theoretical formalism and wrote the manuscript. All authors discussed the results and commented on the manuscript.

**Additional Information**

Supplementary information is available for this paper

Correspondence and requests for materials should be addressed to R.M. (raffaele.mezzenga@hest.ethz.ch).

**Competing Financial Interest** Authors declare no competing financial interest.

Supplementary Information for:

# Flow-induced order-order transitions in amyloid fibril liquid crystalline tactoids

Hamed Almohammadi[1], Massimo Bagnani[1], Raffaele Mezzenga[1,2*]

[1]Department of Health Sciences and Technology, ETH Zurich, Zurich, Switzerland.

[2]Department of Materials, ETH Zurich, Zurich, Switzerland.

*Correspondence to: raffaele.mezzenga@hest.ethz.ch

I. Length and height distribution of amyloid fibrils

Figure S1 presents the length and height distributions of the amyloid fibrils used in this study, obtained by analyzing the AFM images using the FiberApp software developed by Usov and Mezzenga[1]. The lognormal distributions were fitted in MATLAB resulting in fitting parameters of $\mu_{fitting} = 5.5 \pm 0$ and $\sigma_{fitting} = 0.6 \pm 0$ for the length and $\mu_{fitting} = 0.9 \pm 0$ and $\sigma_{fitting} = 0.3 \pm 0$ for the height distributions.

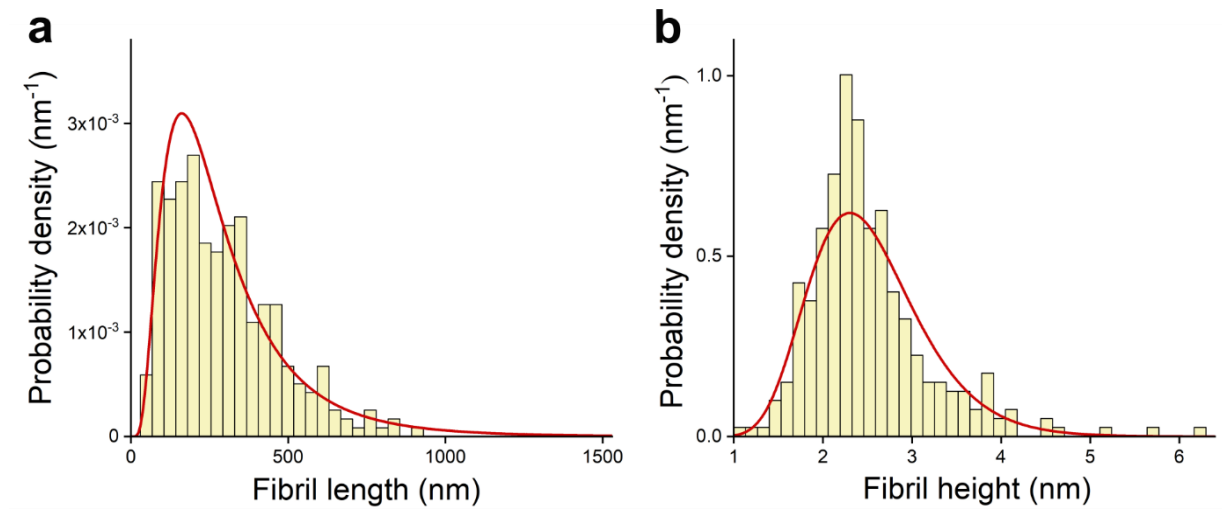

**Fig. S1 | Length and height distribution of amyloid fibrils.** The solid curves are the lognormal distribution fitted to data where $n = 300$. **a,** the arithmetic and weighted mean length values are $L_m = 302.9$ nm and $L_w = 424.5$ nm, respectively. The fitting parameters are: $\mu_{fitting} = 5.5 \pm 0$ and $\sigma_{fitting} = 0.6 \pm 0$. **b,** The mean arithmetic height value is $h_m = 2.5$ nm with fitting parameters $\mu_{fitting} = 0.9 \pm 0$ and $\sigma_{fitting} = 0.3 \pm 0$.

## II.    Determination of order-order transition

Here we provide further details to support deformation-induced order-order transitions between different classes of liquid crystalline tactoids, as shown in Fig. S2. In the main text (Fig. 2), we provided the tactoids deformation when their long axes (or short axis) is placed close to ~45º relative to crossed polarizers during the deformation in contraction zone. Such configuration allows to have long and short axis of the tactoids to be visible for unambiguous measurement and study of deformation. To do so, we simply placed the microfluidic chip in a way that the flow direction is in ~45º with respect to crossed polarizers, as the tactoids long axis align in the flow direction in the contraction zone, Fig. S2a. In this setup, the order-order transition of the bipolar to homogenous tactoids is identified when curvilinear orientation of the director in the bipolar that leads to appearance of the invisible corner for the tactoid changes to uniformly aligned director field that is seen as almost uniform white texture, Fig. S2c. We determined the cholesteric to homogenous transition, when the stripped texture of the tactoids changes to uniform white texture for the tactoid, Fig. S2d.

To further support the order-order transition from bipolar to homogenous tactoid, following the experimental protocol designed in Ref. 2 to distinguish different classes of tactoids by rotating the tactoids under crossed polarizers, we performed experiments in such a way to have the tactoids to be parallel to one of the crossed polarizers. This is achieved by placing the microfluidic channel in such a fashion that the flow direction is parallel to one of the crossed polarizers, Fig. S2b. In this configuration, the bipolar tactoid is determined when a central dark cross is seen, while the homogenous tactoid should be invisible[2]. Thus, using this setup, the transition from bipolar to homogenous tactoid should happen, when the texture with central dark cross becomes invisible. This is clearly seen in Fig. S2e.

Fig. S2f shows the cholesteric to homogenous tactoid transition. Upon elongation of the cholesteric tactoid whose long axis is almost parallel to one of the crossed polarizers (but not exact parallel, as we want to show its stripped texture) and reaching a critical elongation ratio, the stripped texture becomes invisible. This shows order-order transition of cholesteric to homogenous phases, and rules out the possibility of transition to any other class other than homogenous.

It should be noted that although in equilibrium condition, the order of the transition is from homogenous to bipolar and then to cholesteric with an increase in volume and decrease in aspect ratio, here due to non-equilibrium nature of the system bipolar configuration is not observed when a cholesteric tactoid is elongated. To support this argument, we look at the transition from a theoretical perspective and compare the results with the data presented in Figure 4a in the main text. Let's assume that there is a cholesteric, to bipolar and then to homogenous tactoids transition. This implies that there should be a transition from bipolar to homogenous in the aspect ratio that is less than ~17 (see Fig. 4a), as we already see homogenous tactoids for the aspect ratio higher than ~17 when the cholesteric tactoids are deformed. However, following the theory (Eq. 3 in the main text) that predicts bipolar to homogenous tactoids transition, the expected aspect ratio for bipolar to homogenous transition for a tactoid with initial cholesteric configuration is ~56. This prediction is much higher than the range that we see homogenous tactoids and shows the invalidity of the assumption that there is a cholesteric, to bipolar and then to homogenous tactoids transition.

Note that, in Fig S4e and f, after the transition, the homogenous tactoids with director field parallel to one of the crossed polarizers, look darker than the isotropic medium. This is so as statistically the number of the rods that are in parallel to one of the crossed polarizers is higher when one looks though the tactoids in the chip compared to isotropic medium. This is also reflected in Fig. 2 in Ref. 2 when a homogenous tactoid is place in parallel to one of the crossed polarizers. In Fig. S2c and d the inhomogeneous intensity of birefringence along the deformed tactoids is related to the heterogeneity of the thickness of the tactoids.

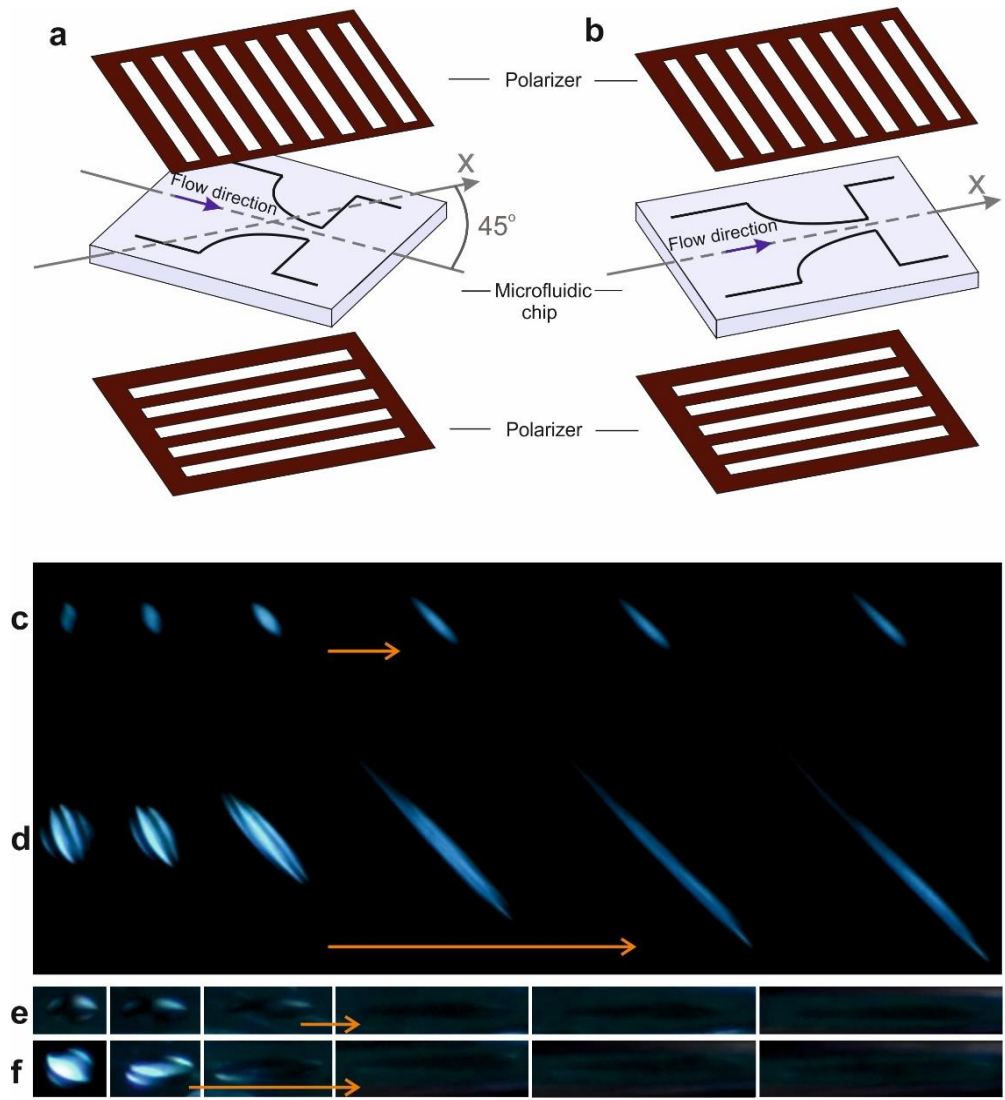

**Fig. S2 | Deformation-induced order-order transitions observed at two different angles with respect to fixed crossed polarizers. a** The microfluidic chip is placed in a way to have 45° between the extension ($\dot{\varepsilon}_{xx}$) direction and crossed polarizers. **b** The microfluidic chip is rotated 45° degree compared to (**a**), resulting in 0° or 90° between extension direction and crossed polarizers. **c-d** The bipolar and cholesteric tactoids are elongated in configuration (**a**) resulting in observing of the elongation of the tactoids in 45 ° with respect to crossed polarizers. The phase transition from bipolar and cholesteric to homogenous is shown with red arrow (the extended red arrow in (**d**) shows the phase that is defined as transition phase). **e-f** The bipolar (**e**) and cholesteric (**f**) tactoids are elongated using configuration depicted in (**b**). The transition to homogenous is determined when the tactoids become invisible.

### III. Determination of the Viscosities

For hydrodynamic analysis in this study, we assume the viscosity of the medium and tactoids to be equal to viscosity of the isotropic and nematic phases, respectively. Thus, viscosity of the isotropic and nematic phases are measured following the description provided in Methods. The desired amount of suspension from isotropic and nematic phases from a suspension of the amyloid fibrils that was already phase separated were taken and analyzed. As shown in Figure S3, the zero shear viscosity of the nematic phase is lower than the isotropic phase, which can be related to alignment of the fibrils in nemaic phase. Note that, although the viscosities of the liquid crystalline phases vary depending on the shear rate, following the common assumption in the context of the droplet deformation[3], we used zero shear viscosity values for hydrodynamic analysis in the main text.

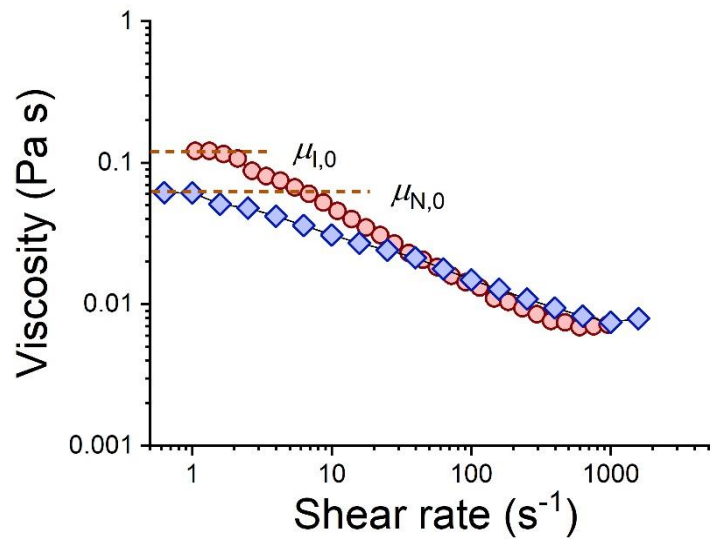

**Fig. S3 | Viscosity of the isotropic and nematic phases of amyloid fibrils versus shear rate.** The dashed line denotes zero shear viscosity that is equal to 0.121 and 0.061 Pa s for isotropic and nematic phases, respectively.

## IV. Oil-in-water control experiments

To show the effects of the low interfacial tension of the tactoids, experiments with simple fluids (oil droplet in water-glycerol mixture), with interfacial tension of order of ~0.01 N m-1, are performed and compared against the tactoid (Fig. 2f or Fig. S4). To be systematic in the comparison, the relevant parameters in the droplet deformation[3], i.e. droplet size, the viscosity of the droplet, viscosity of the medium, and accordingly the viscosity ratio of droplet to medium, as well as extension rate, were kept essentially the same. We used water-glycerol mixture with viscosity of 0.122 Pa s (is taken from Ref. 4) as medium phase and olive oil with viscosity of 0.063 Pa s (measured directly) for droplet phase, having similar viscosity as liquid crystalline system where the viscosity of the surrounding phase and tactoid are, respectively, 0.121 Pa s and 0.061 Pa s (see previous section). The droplet size and extension rate of the olive oil droplet are chosen to be almost equal to the tactoid that is shown in Fig. 2c (or Fig. S4a), i.e. olive oil: droplet radius = 27 µm and extension rate = 0.020 s$^{-1}$, tactoid: equivalent radius of tactoid = $(r^2R)^{1/3}$ = 26 µm and extension rate = 0.017 s$^{-1}$. Comparing Figures 2c and 2f (or Figures S4a and S4b), it is clear that, while tactoid reaches to aspect ratio of ~38 (or the final to initial length ratio of ~10), the oil droplet remains almost undeformed. This clearly illustrates the remarkable effects of the very low interfacial tension of the tactoids.

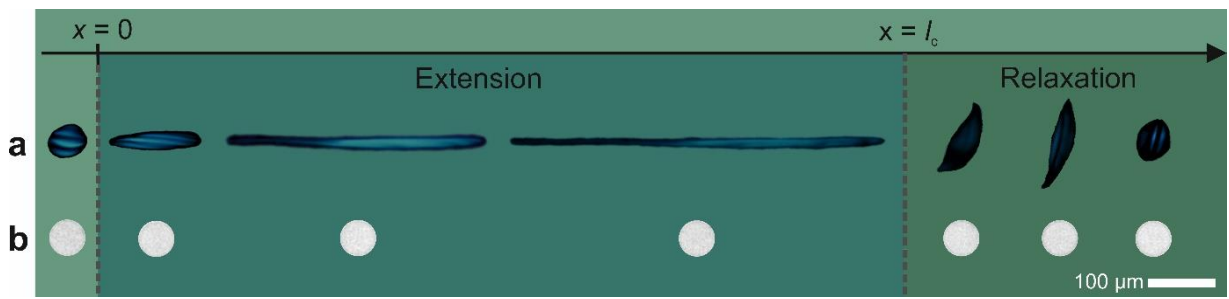

**Fig. S4 | Comparison of deformation of liquid crystalline and olive oil droplet ion water-glycerol solutions.** The extension rates are (a) 0.017 and (b) 0.020 s$^{-1}$. While the tactoid reaches to aspect ratio of ~38 or the final to initial length ratio of ~10 (a), the oil droplet remains almost unreformed (b).

## V. Determination of the density, interfacial tension, anchoring strength, splay and bending elastic constant, and twist elastic constant

**Density.** To get density of continues phase to be used in Reynolds number in the main text, similar to viscosity, we assume the density of the medium to be equal to density of the isotropic phase, $\rho_I$, and write it as a function of amyloid fibrils density, $\rho_{amy}$, and water density, $\rho_{H_2O}$, using:

$$\rho_I = \frac{m_{amy} + m_{H_2O}}{V_{amy} + V_{H_2O}} = \frac{\rho_{amy} V_{amy} + \rho_{H_2O} V_{H_2O}}{V_{amy} + V_{H_2O}} \tag{S1}$$

where $m_{amy}/V_{amy}$ and $m_{H_2O}/V_{H_2O}$ are the mass/volume of the amyloid fibrils and water, respectively. To be able to get $\rho_I$ in Eq. S1, we relate $V_{amy}$ and $V_{H_2O}$ using following relation to isotropic phase concentration, $c_I$:

$$c_I = \frac{m_{amy}}{m_{amy} + m_{H_2O}} = \frac{\rho_{amy} V_{amy}}{\rho_{amy} V_{amy} + \rho_{H_2O} V_{H_2O}} \tag{S2}$$

Substituting $c_I = 0.02$ wt, $\rho_{amy} = 1.3$ g cm$^{-3}$ (taken from Ref. 5), and $\rho_{H_2O} = 1.0$ g cm$^{-3}$ in Eq. 2, we get $V_{amy} = 0.016\, V_{H_2O}$. This together with density values substituted in Eq. S1 results in $\rho_I = 1.005$ g cm$^{-3}$.

**Interfacial tension.** The interfacial tension of the tactoids is estimated using universal scaling law that predicts the interfacial tension of the phase separated colloidal suspension of hard rods as[6-7]:

$$\gamma = b \frac{k_B T}{L_{particle} D_{particle}} \tag{S3}$$

where $k_B$ is Boltzmann constant, $T$ is the temperature, $L_{particle}$ and $D_{particle}$ are the length and diameter of the hard rods, respectively, that are taken to be equal to $L_w$ and $h_m$ of the fibrils here. The term $b$ is a constant that is taken to be 0.3 value[2,7]. Substituting all the parameters, we get interfacial tension to be $\gamma = 1.1 \times 10^{-6}$ N m$^{-1}$.

**Anchoring strength.** To get the anchoring strength, we followed the Ref. 8 showing linear dependence of anchoring strength on fibrils length. Thus, knowing the anchoring strength for fibrils with average length of 401 and 525 nm to be 1.45 and 1.8, respectively, from Ref. 2 and Ref. 8, we estimated the anchoring strength for fibrils with average length of 302.9 nm used in this work to be 1.17. Similar anchoring strength value, 1.06±0.27, is found using Wulff construction $\omega = (\alpha/2)^2$, that is for when anchoring strength is higher than one and $\alpha$ is the aspect ratio of the homogenous tactoids at equilibrium condition[6]. This also confirms the linear dependence of the anchoring strength on fibrils length as discussed in Ref. 8.

**Splay and bending elastic constant, $K$.** The splay and bending elastic constant ($K$) is found by fitting the relation $\alpha = \frac{4K}{\gamma R}$ to the data that shows the aspect ratio of bipolar tactoids as a function their long axes at equilibrium condition, Fig. S5. This derivation that relates the shape of the bipolar tactoids to $K$ is provided in Ref. 2 by minimizing the energy of the bipolar tactoids at equilibrium condition.

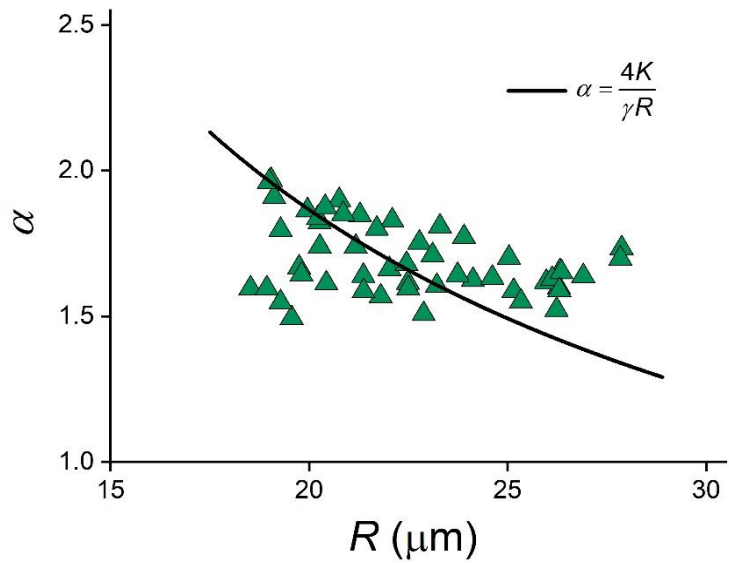

**Fig. S5 | Aspect ratio of the bipolar tactoids as a function of their long axis at equilibrium condition.** The solid curve obeying $\alpha = \frac{4K}{\gamma R}$ indicates the best fit to data with fitting parameter $K = 1.0 \times 10^{-11}$ N.

**Twist elastic constant, $K_2$.** Similar to Ref. 2, we find $K_2$ using[9]:

$$K_2 \sim k_B T L_{\text{particle}}^2 n \tag{S4}$$

where $n$ is the number of rod-like objects. Having the all values for the parameters in Eq. S4 and $K_2$ to be $0.559 \times 10^{-12}$ N for the amyloid fibrils with weighted mean length ($L_{502}$, lets use the length value as a subscript in this section) of 502 nm and concentration ($c_{502}$) of 0.013 g cm$^{-3}$ from Ref. 2. We take these values as a reference and find the $K_2$ for the amyloid fibrils suspension used in this study that is with weighted mean length ($L_{424.5}$) of 424.5 nm and concentration of $c_{424.5} = 0.022$ g cm$^{-3}$, using:

$$K_{2_{424.5}} = \left(\frac{T_{424.5}}{T_{502}}\right)\left(\frac{L_{424.5}}{L_{502}}\right)^2 \left(\frac{n_{424.5}}{n_{502}}\right) K_{2_{502}} \tag{S5}$$

where *n* parameter can be substituted by $c/(\rho_{particle} V_{particle})$ with *c* the concentration, and $\rho_{particle}$ and $V_{particle}$ (~ $D_{particle}^2 L_{particle}$) the density and volume of rod-like objects, respectively. To get the volume of the amyloid fibrils, one need to take the fibrils as a double stranded cylinder. This together with simplification of equation S5 can give:

$$K_{2_{424.5}} = \left(\frac{L_{424.5}}{L_{502}}\right)\left(\frac{c_{424.5}}{c_{502}}\right)\left(\frac{D_{502}}{D_{424.5}}\right)^2 K_{2_{502}} \tag{S6}$$

Now substituting all of the above reported parameters and the values of $D_{502} = 2$ nm and $D_{424.5}=1.25$ nm (these are diameter for protofilament[10]), we get twist constant for our used amyloid fibrils suspension in this study to be $K_{2_{424.5}}= 2.0 \times 10^{-12}$ N.

## VI. The shape of the tactoids in the straight channel

As it is mentioned in the main text, the tactoids need to travel the straight channel with velocity $U$ before entering the contraction zone. Here we compare the aspect ratio of the tactoids as a function of their volume at equilibrium condition (the sample is placed in a cuvette) and in the microfluidic channel before entering to the constriction zone at $x=0$ in Fig. 1 (defined as initial shape in the main text), see Fig. S6. The velocity values reported in Fig. S6 are the flow speed that are tested in this study to get different extension rate in Fig. 3. As it can be seen, the aspect ratios of homogenous tactoids in all tested conditions and bipolar tactoids at low flow velocity ($U=0.5$ µm s$^{-1}$), are in the range of equilibrium state. The aspect ratio of cholesteric and bipolar tactoids are higher than the equilibrium state and around three at the highest flow speed. However, such deviation of the initial shape of the tactoids from equilibrium state when these are compared to the range of aspect ratio that the tactoids experience are very small (3/25=0.12). Thus, it enables to extend the linear deformation argument of the tactoids for this small deviation range as well, as rationalizing by Eq. 1 which is used to predict the deformation of the tactoids from its equilibrium state.

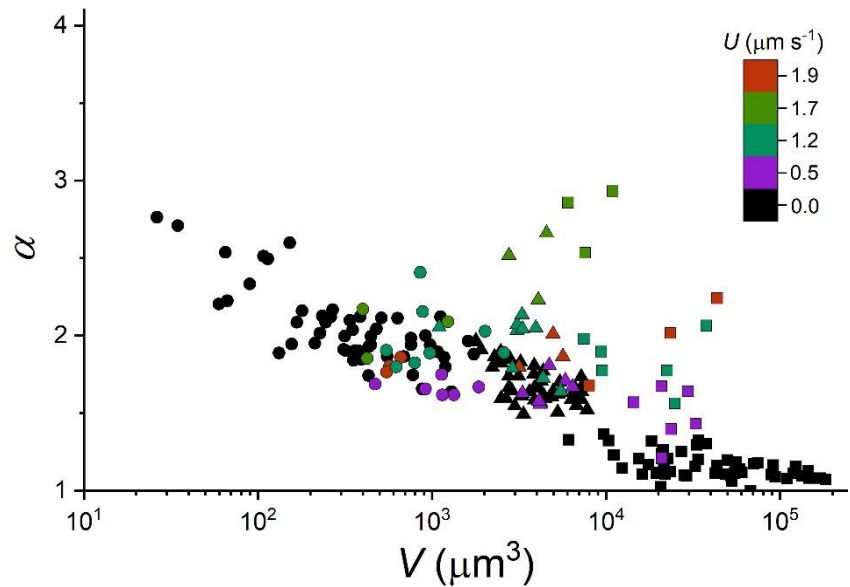

**Fig. S6 | Aspect ratio of the tactoids as a function of their volume at equilibrium condition and in the microfluidic channel moving with velocity $U$.** The velocity ($U$) values are the flow speed that are tested in this study to get different extension rate in Fig. 3. The black symbols ($U = 0$ m s$^{-1}$) refer to the sample that is placed in a cuvette.

## VII. Determination of $P_\infty$

Figure S7 shows the bulk cholesteric phase of the used amyloid fibrils ($L_m$ = 302.9 nm), showing the $P_\infty$ to be 25.5 μm. This, together with the $P_\infty$=20 μm for fibrils with $L_m$ = 401 nm and $P_\infty$=15 μm for fibrils with $L_m$ = 525 nm in Ref. 2 and 8, further supports the argument in Ref. 8 pointing at an increase in the pitch of the cholesteric phase with a decrease in fibrils length.

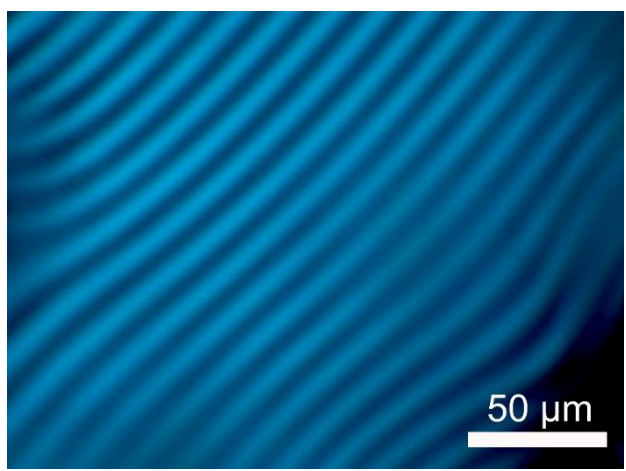

**Fig. S7 | Cholesteric bulk phase of amyloid fibrils solution.** The value of $P_\infty$ is found to be 25.5 μm.